 \documentclass[doublespacing]{elsart}
 \usepackage{graphics}
\usepackage{amssymb}
\begin{document}

\begin{frontmatter}


 %\title{Title\thanksref{label1}}

\title{Sensitivity and spatial resolution of square loop SQUID magnetometers}

%\corauth[cor]{S Rombetto}
\author [CNR]{S Rombetto \ead{s.rombetto@cib.na.cnr.it}}
\author[CNR]{A  Vettoliere}
\author[CNR]{C Granata}
\author[CNR]{M Russo}
\author[CNR]{C Nappi}
\address[CNR]{Istituto di Cibernetica ``E. Caianiello" del CNR, Via Campi Flegrei 34, 80078, Pozzuoli (Naples), Italy}

\begin{abstract}
% Text of abstract
We calculate the flux threading the pick-up coil of a square SQUID
magnetometer in the presence of a current dipole source. The result
reproduces that of a circle coil magnetometer calculated by Wikswo
\cite{Wikswo_78} with only small differences. However it has a
simpler form so that it is possible to derive from it closed form
expressions for the current dipole sensitivity and the spatial
resolution. The results are useful to assess the overall performance
of the device and to compare different designs.
\end{abstract}

\begin{keyword}
magnetometers \sep spatial resolution \sep dipole sensitivity \sep
current dipole
\PACS 87.85.Ox, 87.85.Tu, 74.90.+n
\end{keyword}
\end{frontmatter}

% main text
\section{Introduction}
%\label{}
The application of Superconducting QUantum Interference Devices
(SQUIDs) to the measurement of biomagnetic fields  has occurred
because of their sensitivity, their stability and their flexibility.
In fact SQUID based devices offer the possibility to implement
measurements where no other methodology is possible and moreover
they present the advantage to be a non-invasive technique
\cite{Fagaly}, \cite{SternickelBraginski}. Usually studies, aimed to
develop better devices, focus on gradiometric configurations since
they are less sensitive to the noise. Hereafter we focus on
magnetometer configuration which is successfully employed in
multichannel systems for biomagnetic imaging \cite{Chieti2}.
Magnetometers are extremely sensitive to the outside environment,
while some other configuration, like gradiometers provide the
advantage of discriminating against unwanted background fields from
distant sources while retaining sensitivity to the
nearby sources.\\
In a dc-SQUID magnetometer, the pick-up coil, collects the magnetic
flux giving an effective area much larger than that of the SQUID
itself \cite{Carmine_2001}. Here we study the effect of magnetometer
pick-up coil geometry on the performances of SQUID devices for
biomagnetism. Well known and widely spread expressions for the flux
threading a magnetometer and the current dipole sensitivity have
been calculated by Wikswo \cite{Wikswo_78} referring to a circular
loop device. Since typically, SQUID magnetometers present a square
pick-up loop \cite{Carmine_2001}, we were driven,  for the best
characterization of such devices, but also for general reasons, to
recalculate  the quantities of interest in the case of square
pick-up loop.  It turned out that expressions for the minimum
detectable current dipole and for the spatial resolution are easily
derived for the square geometry.
\section{Magnetic flux threading a square magnetometer in the current dipole model}
\begin{figure}[h!]
\includegraphics{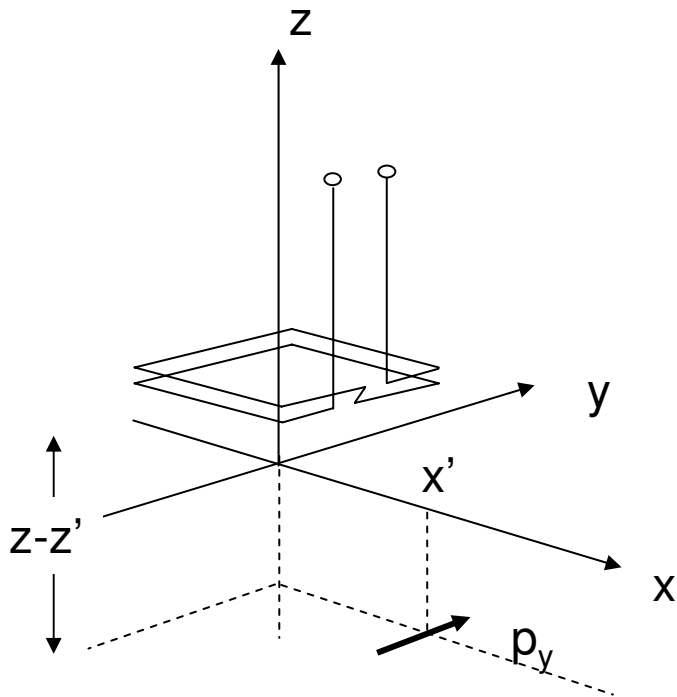}
\caption{\emph{Schematic of the square loop magnetometer.  A current dipole source $\vec{p}$  is placed at the
point $\vec{r'}=(x',0,z')$.}}\label{configuration}
\end{figure}

A widely used mathematical model to describe bioelectric currents is
the current dipole \cite{Chieti1}. It is a good model of elementary
cellular events, thus it can be used  for
magnetoencephalography as well for magnetocardiography studies.\\
Let us consider a dipolar electric current source
$\vec{p}\left(p_x,p_y\right)$ located at a point $\vec{r'}=(x',0
,z')$ in a conducting half space and a pick-up loop centered on the
z-axis (Fig.\ref{configuration}). The magnetic field generated by
the source $\vec{p}$ at a point $\vec{r}=(x,y,z)$ has the vector
potential $\overrightarrow{A}({A_x,A_y,0})$
\begin{equation}\label{potential}
A_{x,y}=\frac{\mu_0 p_{x,y}}{4 \pi
\sqrt{\left(x-x'\right)^2+\left(y-y'\right)^2+\left(z-z'\right)^2}}
\end{equation}
By using eq.(\ref{potential}) it is possible to calculate the
magnetic flux through the considered pick-up coil by performing a
line integral around the loop. In the case of a circular loop having
radius $R$, Wikswo \cite{Wikswo_78} derived the  result
\begin{equation}\label{PhiCircular}
\Phi=\frac{\mu_0 p_y}{k
\pi}\sqrt{\frac{R}{x'}}\left[\left(1-\frac{k^2}{2}\right)K(k)-E(k)\right]
\end{equation}
where
\begin{equation}
k^2=\frac{4x'R}{\left[(x'+R)^2+(z-z')^2\right]}
\end{equation}
$K(k)$ and $E(k)$ are respectively the complete elliptic integrals
of first and second kind. In a very similar way we have calculated
the flux collected by a square loop magnetometer having size $L$,
laying parallel to the $xy$ plane, at a distance $D=z-z'$ above a
current dipole source $\vec{p}$ (Fig.\ref{configuration}). The
result is
\begin{equation}\label{Phi_square}
\Phi=\frac{\mu_{0}p_y}{2\pi}\left[\sinh^{-1}\left(\frac{L}
{\sqrt{4D^2+(L-2x')^2}}\right)-\sinh^{-1}\left(\frac{L}{\sqrt{4D^2+(L+2x')^2}}\right)\right]
\end{equation}
Note that $p_x$ does not contribute to the collected flux for
symmetry reasons, so that there is no loss of generality if we
consider $\vec{p}$ as having only the $y-$component. In order to
compare devices presenting different geometries (square loop or
circle loop),  we shall consider equal area devices, that is
equivalent to the condition $R=L/\sqrt{\pi}$. Also we shall
introduce in the above equations the dimensionless source position
$\overline{x}=x'/D$, the reduced flux $\Phi\pi/\mu_0p_y$ and  the
geometrical parameter $q=L/D$. As we shall see, although
eq.(\ref{PhiCircular}) and eq.(\ref{Phi_square}) give, almost, the
same result, eq.(\ref{Phi_square}) allows for useful analytical
progresses which eq.(\ref{PhiCircular}) does not permit.
\section{Maximum magnetic flux}
\begin{figure}[h!]
\includegraphics{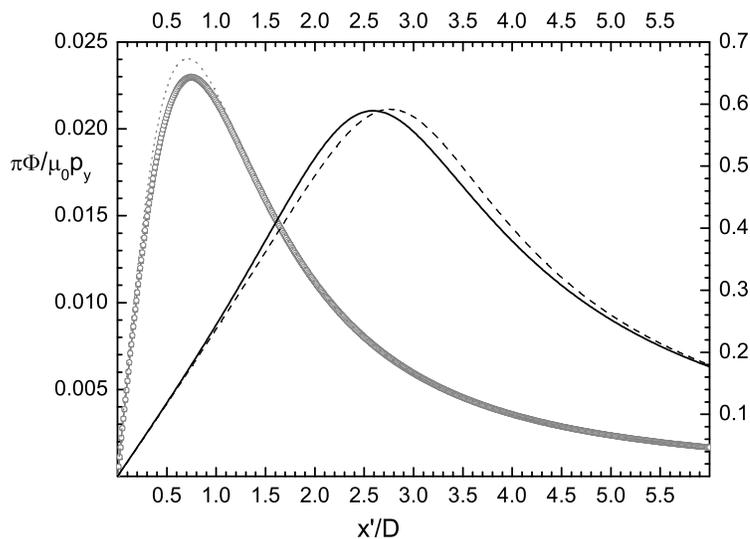}\caption{\emph{Comparison
between the normalized flux threading devices presenting square
or circle shape  as a function of the source position
$x'/D$. Gray curves on the left have been obtained using $q=0.5$ in eq.(\ref{Phi_square}).
The two black curves on
the right have been obtained using  $q=5$ in eq.(\ref{Phi_square}). The gray dotted curve
shown on the left has been obtained by using for both geometries
the Taylor expansion, given by
eq.(\ref{expansion}).} \label{flux_tutti}}
\end{figure}

In order to evaluate the minimum detectable current dipole and the
spatial resolution as a function of the geometrical parameters $L$
and $D$, it is essential to determine the value of the maximum flux
and the
maximizing source position $x'=x'_{max}$ for any $q$.\\
In Fig.\ref{flux_tutti} the reduced flux, calculated by means of
eqs.(\ref{PhiCircular}), (\ref{Phi_square}), is plotted as a
function of the reduced source position $\overline{x}$ for two
different values of the ratio $q$,  for the two geometries, circle
and square. As a general picture one sees that the flux is zero when
the source is exactly under the loop ($x'=0$) and, as the source
moves away, it maximizes for $x'=x'_{max}$, before decaying
out.\\
While the elementary procedure (zeros of the derivatives) for the
maximum finding does not give straightforward results for
eq.(\ref{PhiCircular}) and eq.(\ref{Phi_square}),  simple
approximate analytical results can be obtained for the maximum flux,
on the basis of physical considerations. First consider the case of
a large loop size to source distance ratio ($q\gg 1$). It is evident
that in this case (in exact manner in the limit of zero distance)
the collected flux  is maximum when the source position coincides
with the loop edge. This is to say $x'=L/2$ ($\overline{x}=q/2$) and
$D\rightarrow0$ for a square loop, and analogously, $x'=R$
($\overline{x}=q/\sqrt{\pi}$) and $D\rightarrow0$ for a circle loop.
Thus we can take this asymptotic values, $q/2$ and $q/\sqrt{\pi}$,
as approximate maximum positions for the circle and the square loop,
as it is shown in Fig.\ref{flux_tutti}, where the two maxima
correspond roughly to the points $\overline{x}=2.5$ and
$\overline{x}=2.82$ respectively, since q=5 for the curves in
Fig.(\ref{flux_tutti}).\\In the case of a square loop, using the
value $q/2$ in eq.(\ref{Phi_square}), we are lead to the following
result for the maximum flux
\begin{equation}\label{phimax1}
\Phi_{max}^{square}=\frac{\mu_0 p_y}{2 \pi} \left[
\sinh^{-1}\left(\frac{q}{2}\right)-
\sinh^{-1}\left(\frac{q}{\sqrt{4+4q^2}}\right)\right]
\end{equation}
Now we turn to the case when the ratio between the loop size and the
distance is quite small ($q\ll 1$). Equations (\ref{PhiCircular})
and (\ref{Phi_square}) give in practice identical results for the
two loop geometries and the difference between the fluxes threading
devices with different shape cannot be appreciated: the two results
overlap.\\
For large distances above the source, or small loop area, equations
(\ref{PhiCircular}),(\ref{Phi_square}) can be simplified by
substituting $R=L/\sqrt{\pi}$ in eq.(\ref{PhiCircular}), and then
expanding these expressions in the small parameter $q$. Both
equations give the same result
\begin{equation}\label{expansion}
\Phi=\frac{{\mu_{0}p_{y}}}{4}\frac{\overline{x}}{\left(1+\overline{x}^2\right)^{3/2} }\left(\frac{L}{D}\right)^2=\frac{{\mu_{0}p_{y}}}{4\pi}\frac{\overline{x}}{\left(1+\overline{x}^2\right)^{3/2} }\left(\frac{R}{D}\right)^2
\end{equation}
which is shown in Fig.\ref{flux_tutti} by the gray curves on
the left, calculated for $q=0.5$.\\
\\
thus eq.(\ref{expansion}) represents the flux as a function of
$\overline{x}$, threading a magnetometer positioned at large
distance, with respect to the loop size, above the source. In
Fig.\ref{flux_tutti} the dotted curve represents results obtained by
eq.(\ref{expansion}). In this regime ($q\ll 1$) circle and square
loop, give identical results, so that any information about the
shape of the magnetometer loop is lost.  It is easily found that
eq.(\ref{expansion}) maximizes exactly for
$\overline{x}=\frac{\sqrt{2}}{2},(x'=\frac{\sqrt{2}}{2}D)$ so that
 the maximum flux, for  small loop size to source distance ratio ($q
\ll 1$) is
\begin{equation}\label{max_appr}
\Phi_{max}=\left(\frac{L}{D}\right)^{2}\frac{\mu_{0}p_{y}}{6\sqrt{3}}=\left(\frac{R}{D}\right)^{2}\frac{\pi\mu_{0}p_{y}}{6\sqrt{3}}
\end{equation}
Thus small differences in the collected flux, due to the
inhomogeneity of the source, emerge between square and circle
geometry only when the source is close faced to the loop. This
situation is illustrated in Fig.\ref{flux_tutti} by the black curves
on the right, calculated for $q=5$. As can be seen, the two curves
maximize
in slightly different points, as already observed.\\
A quantitative evaluation, based on numerics, of the validity of the
approximations given by eq.(\ref{phimax1}) and eq.(\ref{max_appr})
as well as of the maximizing positions  will be given in the next
section. We close this section by an estimation of the maximum flux
for a typical situation. If we consider for the current dipole the
value $p_y=10 nA\cdot m$ we find that $\Phi^{square}_{max}=9.18\cdot
10^{-17}$ Wb $\cong45$ m$\Phi_0$ for $q=0.5$ (obtained by using the
values $L=9$ mm, $D=1.8$ cm).
\section{Minimum detectable current dipole }
In order to determine the smallest detectable current dipole
$p_y^{min}$, we have to impose that the collected flux
$\Phi_{max}^{square}$ is comparable to the total flux noise
$\Phi^*$. Therefore $p_y^{min}$ is an evaluation for the sensitivity
of the considered device: if a device can detect a smaller $p_y$,
then it presents a higher sensitivity.\\For large $q$ values ($q\gg
1$), from eq.(\ref{phimax1}) for $\overline{x}=q/2$ we obtain
\begin{equation}\label{sensitivity}
p_y^{min}=\frac{2 \pi \Phi^*}{\mu_0} \left[
\sinh^{-1}\left(\frac{q}{2}\right)-
\sinh^{-1}\left(\frac{q}{\sqrt{4+4q^2}}\right)\right]^{-1}
\end{equation}
For  small $q$, from eq.(\ref{max_appr}) we obtain
\begin{equation}\label{sensitivity2}
p_y^{min}=\frac{6\sqrt{3}\Phi^*}{\mu_{0}q^2}
\end{equation}
The dependence of the sensitivity on the loop to source distance
ratio described by eqs.(\ref{sensitivity}), (\ref{sensitivity2}) is
shown in Fig.$\ref{RS}$. In the same figure it is also shown for a
comparison, $p_y^{min}$ evaluated by a numerical maximum finding
procedure directly from eq.(\ref{Phi_square}). \\For very small
 loop size to source distance ratio ($q->0$) the current dipole sensitivity diverges
as $q^{-2}$, due to the small area of the SQUID pick-up loop. In the
opposite limit, i.e. for very small distance between source and
sensor or very large SQUID sensors ($q->\infty$), the sensitivity
improves without limits ($p_{y}->\infty$) because the collected flux
continues to grow.
\section{Spatial resolution}
When the current dipole source  moves from the position of the
maximum flux along the $x$ direction, with a displacement $\delta$,
there is a change in the flux $\Delta\Phi$. In a general way, for
small $\delta$, one obtains the following expression
\begin{eqnarray}\label{def_risspaz1}
\Delta\Phi=\Phi(x')|_{x'_{max}+\delta}-\Phi(x')|_{x'_{max}}=\nonumber\\
\left(\Phi(x')|_{x'_{max}}+\frac{\Phi''}{2}|_{x'_{max}}\delta^2\right)-\Phi(x')|_{x'_{max}}=\frac{\Phi''}{2}|_{x'_{max}}\delta^2
\end{eqnarray}
where $x'_{max}$ is the value for which the flux maximizes and the
condition $\Phi'(x'_{max})=0$ has been used.
\\If we now assume that the spatial resolution is ``the least
detectable displacement" corresponding to a variation in flux equal
to the flux noise $\Phi^*$, by inverting eq.(\ref{def_risspaz1}) one
obtains
\begin{equation}\label{Delta}
\delta^2=\frac{2\Phi^*}{|\Phi''(x'_{max})|}
\end{equation}
The smaller is $\delta$, the better is the resolution.\\
\begin{figure}[h!]
\includegraphics{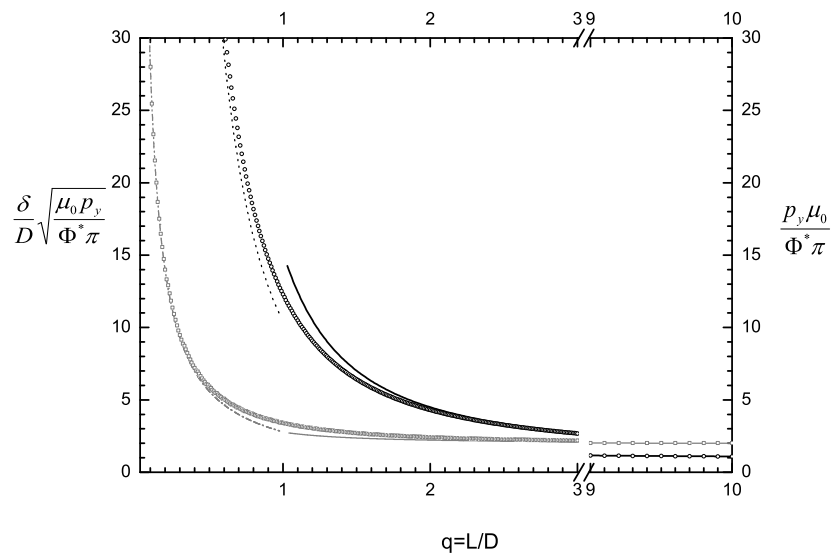}\caption{\emph{The spatial resolution and the current dipole
sensitivity versus the ratio of the side of a square sensor to the
source-sensor distance. Black curves describe
the  current dipole
sensitivity: solid curve is for the case $q\gg 1$ (eq.\ref{sensitivity}), dotted curve is for
the case $q\ll 1$ (eq.\ref{sensitivity}) and circles are for the numerical calculation. Gray curves describe the spatial resolution: solid curve is for
the case $q\gg 1$ (eq.\ref{RSygrande}), dashed curve is for
the case $q\ll 1$ (eq.\ref{RSypiccolo}) and squares are for the numerical calculation. }}\label{RS}
\end{figure}
We now derive an analytical expression for the spatial resolution.
For large loop size to source distance ratio ($q\gg 1$), as we have
seen before,  the flux maximizes approximatively when the condition
$x'_{max}(q)=q/2$ is satisfied. In this regime, the expression
derived for spatial resolution from eqs.(\ref{Phi_square}),
(\ref{Delta}), developed around $\overline{x}=\frac{q}{2}$, becomes
\begin{equation}\label{RSygrande}
\overline{\delta}=\frac{\delta}{D}= \frac{\sqrt{\frac{\Phi^*\pi}{\mu_{0}p_{y}}}}{\sqrt{q\left(\frac{1}{4\sqrt{\left(4+q^2\right)}}-\frac{1-\frac{3}{4}q^2-\frac{9}{8}q^4}{\left(1+q^2\right)^2\left(4+5q^2\right)^{3/2}}\right)}}
\end{equation}
For  small loop to source distance ratio  ($q\ll 1$),  the
analytical expression for the spatial resolution can be obtained on
the basis of eqs.(\ref{expansion}) and (\ref{Delta}), for
$\overline{x}=\frac{\sqrt{2}}{2}$, and the expression for spatial
resolution is
\begin{equation}\label{RSypiccolo}
\overline{\delta}=\frac{\delta}{D}=\frac{3}{q}\sqrt{\frac{\sqrt{3}}{2}\frac{\Phi^*\pi}{\mu_{0}p_{y}}}
\end{equation}
When the $q$ value is about 1, the approximations introduced till
for $\overline{\delta}$ now begin to fail, so that it is necessary
to compute the spatial resolution numerically. In order to do this
and for a comparison with the analytical results, we have calculated
analytically the second derivative of the flux  given in
eq.(\ref{Phi_square}) and calculated its value in the  maximizing
position $x'_{max}$ evaluated numerically  for any $q$.
\\In Fig.\ref{RS} the two analytical solutions eq.(\ref{RSygrande}) and eq.(\ref{RSypiccolo}),and the numerical result for the
spatial resolution are plotted in gray. Solid curve is for the case
$q\gg 1$, dashed curve is for the case $q\ll 1$ and squares are for
the numerical calculation. \\It is worth noting that the spatial
resolution $\overline{\delta}$ defined in eq.(\ref{RSygrande}) has
the lower limit (obtained for $q\rightarrow\infty$)
\begin{equation}\label{limitdelta}
\delta=2D\sqrt{\frac{\Phi^*\pi}{\mu_{0}p_{y}}}
\end{equation}
This means that even if we design a device that could collect a very
large flux, the spatial resolution cannot enhance.
\section{Conclusions}
We found expressions for the dipole sensitivity and for the spatial
resolution of a square loop magnetometer starting from
eq.(\ref{Phi_square}) which gives the flux threading the loop, due a
current dipole source.
 Both quantities  show a monotonic  dependence
on the sensor size for a fixed sensor to source distance. The dipole
sensitivity is limited only by the loop size $L$. On the contrary
the spatial resolution has a lower limit given by
eq.(\ref{limitdelta}), meaning that there is no way to improve the
spatial resolution even using a very large loop size device. Thus
for all practical needs the calculations here presented indicate
that when the distance $D$ is comparable with the size of the loop
$L$, the limit spatial resolution is already obtained.
\bibliographystyle{unsrt}
\bibliography{squaremagnetometers_30marzo}
\end{document}